\documentstyle[eqsecnum,prd,aps,epsf]{revtex}

\draft
\begin{document}

\newcommand{\beq}{\begin{equation}}
\newcommand{\eeq}{\end{equation}}
\newcommand{\beqn}{\begin{eqnarray}}
\newcommand{\eeqn}{\end{eqnarray}}
\newcommand{\pa}{\partial}
\newcommand{\vp}{\varphi}
\newcommand{\ep}{\epsilon}
\def\zero{\hbox{$_{(0)}$}}

\twocolumn[\hsize\textwidth\columnwidth\hsize\csname
@twocolumnfalse\endcsname

\begin{center}
{\large\bf{Black hole formation in the Friedmann universe: \\
Formulation and computation in numerical relativity
}}
~\\
~\\
Masaru Shibata$^{1,2}$ and Misao Sasaki$^{1,3}$ \\
$^1${\em Department of Earth and Space Science,~Graduate School of
  Science,~Osaka University,\\ Toyonaka, Osaka 560-0043, Japan
}\\
$^2${\em Department of Physics, University of Illinois at 
Urbana-Champaign, Urbana, Il 61801
}\\
$^3${\em Department of Physics, School of Science, University of Tokyo,
Bunkyoku, Tokyo 113-0033, Japan
}\\
\end{center}

\begin{abstract}
We study formation of black holes in the Friedmann universe. 
We present a formulation of the Einstein equations under the constant
mean curvature time-slicing condition.
Our formalism not only gives us the analytic solution of the perturbation
equations for non-linear density and metric fluctuations on superhorizon 
scales, but also allows us to carry out a numerical relativity
simulation for black hole formation after the scale of the density
fluctuations is well within the Hubble horizon scale.
We perform a numerical simulation of spherically symmetric  
black hole formation in the radiation-dominated, spatially flat 
background universe for a realistic initial condition supplied from the
analytic solution. 
It is found that the initial metric perturbation has to be non-linear 
(the maximum value of 3D conformal factor $\psi_0$ at 
$t=0$ should be larger than $\sim 1.4$) for a black hole to be formed,
but the threshold amplitude for black hole formation and the final black 
hole mass considerably depend on the initial density (or metric) profile 
of the perturbation: The threshold value of $\psi_0$ at $t=0$ 
for formation of a black hole is smaller for 
a high density peak surrounded by a low density region than for 
that surrounded by the average density region of the flat universe. 
This suggests that it is necessary to take into account the
spatial correlation of density fluctuations in the study of
primordial black hole formation. 
\end{abstract}
\pacs{PACS: 04.25.Dm, 95.35.+d, 97.60.Lf}
\vskip2pc]

\section{Introduction}

The formation of black holes in the early universe and its cosmological
implications have been discussed in a variety of contexts for decades
\cite{review}. 
However, it has long been thought that it would be practically
impossible to prove or disprove the existence of these primordial black
holes.
Recent discoveries of microlensing events by MACHOs of mass 
$\sim 0.5 M_\odot$ in the halo of our galaxy\cite{macho} have
dramatically changed this situation. By various other means 
of observation, it may be possible to deny all the other possibilities
and hence to identify MACHOs with black holes. Then these MACHO black
holes must be primordial since it is impossible to form a black hole 
of mass smaller than $\sim M_\odot$ as a result of stellar evolution 
\cite{kippen}. 
Furthermore, it has been recently suggested that if MACHOs are in fact
primordial black holes, the number of binaries that are just coalescing
today may be large enough to be directly detected by the oncoming
gravitational wave observatories such as LIGO, VIRGO, GEO and TAMA 
within a few years\cite{NSTT}.
Consequently, it has become an urgent issue to quantify how and
when these black holes could be formed in a precise manner.

Among other possibilities, primordial black holes are most conceivably
formed from large curvature perturbations generated during an 
inflationary stage of the very early universe \cite{BeLiWa,yokoyama}. 
The curvature perturbations generated in the inflationary universe are 
dominated by the so-called growing adiabatic mode of density
perturbations. In the linear theory, the evolution of these
perturbations is well studied and their temporal
behavior is known throughout the whole stage from the epoch when their
wavelengths are much larger than the Hubble horizon scale until 
their evolution becomes non-linear on scales much smaller than the
Hubble horizon.

However, the amplitude must be already large enough (of order unity) 
to form black holes when the characteristic wavelengths of the
perturbations were on superhorizon scales. Furthermore, the formation of 
a black hole is itself a fully general relativistic phenomenon.
The evolution of non-linear density perturbations on superhorizon scales
was investigated by several authors and the threshold amplitude of
curvature perturbations on superhorizon scales for forming black holes
was estimated\cite{carr,ivanov}.
These previous estimates of the threshold
amplitude were based on approximate analytical treatments and/or on
rather naive numerical simulations, hence are admittedly
crude. Furthermore, there is a crucial reason that requires us an 
accurate estimate as follows: 
According to the inflationary scenario, the probability distribution
of the curvature perturbations is essentially Gaussian and primordial
black holes are produced from the high amplitude tail of the
distribution. Therefore, a small error in the estimate of the threshold
amplitude will result in a large error in that of the number of produced 
black holes. Thus threshold amplitude must be estimated 
accurately.

As a first step to accomplish this purpose, in this paper, 
we present a new formalism by which it is possible to follow the
formation of a primordial black hole throughout the whole stage starting
from the very early universe when the perturbation is well outside the
Hubble horizon to the final stage when a black hole is formed.
More specifically, our formalism not only gives us the analytic solution
of non-linear curvature perturbations on superhorizon scales but also
allows us to perform a numerical simulation of the black hole formation
with the initial data given by the analytic solution, with no need of
changing the gauge conditions or of numerical matching. 
In addition, it may be worthwhile to mention that the constant mean
curvature time-slicing employed here is equivalent to the so-called
constant Hubble slicing in cosmological perturbation theory 
\cite{KS}. And, it has
been pointed out that the constant Hubble slicing is most appropriate
for evaluating non-linear curvature fluctuations generated during
inflation\cite{SasTan}.
Hence the initial curvature perturbation spectrum evaluated in models of
inflation can be directly used for the initial data of our problem.

Then using our formalism, we carry out a spherically symmetric
simulation of black hole formation in the radiation-dominated Friedmann
universe. We consider the initial data with two 
parameters; one describes the amplitude and the other the radial
profile. We find that both the threshold amplitude for black hole
formation and the final black hole mass depend appreciably on the
initial profile of the perturbation. We also consider another possible
criterion for black hole formation by defining a compaction function
of the perturbation profile. Although this function can be defined only
for a spherically symmetric configuration, we find the maximum value of
this function gives us a better criterion for the formation of black
holes.

While this paper was in preparation, a couple of papers on the
primordial black hole formation appeared on the astro-ph\cite{NieJed}.  
It seems that their formalism is powerful for studying the 
formation of a black hole in a spherical symmetric spacetime. 
However, it does not seem convenient 
to give a realistic initial condition which should be supplied 
just after inflation. Actually, they give initial conditions 
at the epoch when the scale of the density fluctuation is 
as small as the Hubble horizon scale. Since the density fluctuation
is already nonlinear at that epoch, it is impossible to control
the initial date so that it reduces to the growing adiabatic mode
when the evolution is traced back in time to the very early universe.
In other words, their initial conditions are inevitably contaminated by
unrealistic decaying mode perturbations which badly diverge as $t\to 0$. 
As a result, though the criterion for black hole formation they find is
new and interesting, it cannot be directly related to the initial
condition at the end of inflation.
So that it is not convenient for a practical study of 
primordial black hole formation.
Furthermore, application of their formalism is restricted 
to the spherical symmetric case (i.e., very special case). 
In contrast, in ours, it is easy to relate a 
criterion of black hole formation to an initial condition 
just after inflation, and also it can 
be applied to general 3D cases. The only restriction of our formalism is
that the spacetime be asymptotically spatially flat Friedmann.

The paper is organized as follows. 
In Sec. II, we present the Einstein and hydrodynamic equations 
in the Friedmann universe using the 3+1 formalism, which have 
appropriate forms for numerical relativity simulations. 
We then introduce the constant mean 
curvature time-slicing in which the equations 
for geometric variables have similar forms to those in 
the maximal slice condition in the asymptotically flat spacetime. 
In Sec. III, assuming that the length scale of a density fluctuation is
always much longer than the Hubble horizon scale, 
we take the long wavelength limit of the 
equations derived in Sec. II, and then find the analytic solution 
for the perturbation equations. 
In Sec. IV, we perform numerical simulations of black hole formation 
in a spherically symmetric, radiation-dominated universe 
using initial conditions given by the analytic solution in Sec. III. 
Sec. V is devoted to summary. 
Throughout this paper, we use the units $c=1=G$. 

\section{Formulation}

We present the Einstein and 
hydrodynamic equations in the Friedmann universe 
using the 3+1 formalism in general relativity. 
We write the line element as 
\beqn
ds^2&=&g_{\mu\nu}dx^{\mu}dx^{\nu} \nonumber \\
&=&(-\alpha^2+\beta_k\beta^k)dt^2
+2\beta_i dx^i dt+\gamma_{ij}dx^i dx^j ,
\eeqn
where $g_{\mu\nu}$, $\alpha$, $\beta^i~(\beta_i=\gamma_{ij}\beta^j)$, 
and $\gamma_{ij}$ are the 4D metric, 
lapse function, shift vector, and 3D spatial metric, respectively. 
Since we consider an asymptotically spatially flat Friedmann universe, 
we rewrite $\gamma_{ij}$ as  
\beq
\gamma_{ij}=\psi^4 a(t)^2 \tilde \gamma_{ij},
\eeq
and impose the condition ${\rm det} (\tilde \gamma_{ij})=
{\rm det}(\eta_{ij})\equiv \eta $, where $\eta_{ij}$ is a flat spatial metric. 
Here, $a(t)$ is defined to be 
the scale factor in the homogeneous universe, i.e., 
the scale factor in the asymptotic region, 
and we determine it from the well known equations for the 
scale factor as 
\beqn
&& \ddot a = -{4\pi \over 3}a \Bigl[\rho_0(t) +3P_0(t)\Bigr],\\
&& \dot a^2={8\pi \over 3}a^2 \rho_0(t),\label{eqsc}
\eeqn
where $\rho_0$ and $P_0$ are the density and 
pressure for the homogeneous universe, and 
$\dot a = \pa_t a$. 
As we find below, $a_0$, $\rho_0$ and $P_0$ are 
automatically determined when an equation of state is provided. 

We also rewrite the extrinsic curvature 
$K_{ij}$ as 
\beq
K_{ij}=\psi^4 a^2 \tilde A_{ij}+{\gamma_{ij} \over 3} K
\eeq
where $K=K_k^{~k}$ and hence $\tilde A_{ij}$ is 
defined to be traceless. The indices of 
$\tilde A_{ij}$ and $\tilde A^{ij}$ are to be 
raised or lowered in terms of $\tilde \gamma^{ij}$ and 
$\tilde \gamma_{ij}$. In numerical computation, we will solve 
$\tilde \gamma_{ij}$, $\tilde A_{ij}$, $\psi$ and $K$  
instead of $\gamma_{ij}$ and $K_{ij}$. 
Hereafter, we use $D_i$ and $\tilde D_i$ as 
the covariant 
derivatives with respect to $\gamma_{ij}$ and 
$\tilde \gamma_{ij}$, respectively. 

As a source of the energy momentum tensor, 
we consider a perfect fluid for 
which the energy momentum tensor is written as
\beq
T_{\mu\nu}=(\rho + P) u_{\mu} u_{\nu} + P g_{\mu\nu},
\eeq
where $\rho$, $P$ and $u^{\mu}$ are 
the energy density, pressure and four velocity, respectively.
Hereafter, we assume a relativistic polytropic equation of state, 
$P=(\Gamma-1) \rho$, where $\Gamma$ is 
a polytropic constant and we assume 
$\Gamma > 1$ in the following discussion. 
For the relativistic polytropic equation of state, 
we get the following relations from Eqs. (\ref{eqsc}) 
\beq
a=a_ft^{2/3\Gamma}~{\rm and}~~\rho_0={1 \over 6\pi\Gamma^2 t^2},
\eeq
where $a_f$ is a constant. 

The hydrodynamic equations are written in the form 
\beqn
&& \pa_t (w \psi^6 a^3 \rho^{1/\Gamma})
+{1 \over \eta^{1/2}}
\pa_{k} ( \eta^{1/2} w \psi^6 a^3 \rho^{1/\Gamma} v^k)=0,
\\
&& \pa_{t} \{w \psi^6 a^3 (\rho +P) u_{j}\}
+{1 \over \eta^{1/2}}
\pa_{k} \{ \eta^{1/2} w \psi^6 a^3 (\rho +P) v^{k}u_{j}\}
\nonumber \\
&&~~~= -\alpha \psi^6 a^3 \pa_j P 
+w \psi^6 a^3(\rho+P)\Bigl\{
-\alpha u^0 \pa_j \alpha  \nonumber \\ 
&& \hskip 3.5cm + u_k \pa_j \beta^k - 
{u_k u_l \over 2u^0} \pa_j \gamma^{kl} \Bigr\},
\eeqn
where $w \equiv \alpha u^0$, and 
\beq
v^k \equiv {u^k \over u^0}=-\beta^k + \tilde \gamma^{kl}
{u_l\over \psi^4 a^2 u^0}.
\eeq

Evolution equations for geometric variables are written as follows 
\cite{gw3d}: 
\beqn
&&(\pa_t - \beta^k \pa_k) \tilde \gamma_{ij} 
=-2\alpha \tilde A_{ij}  \nonumber \\
&& \hskip 2.cm 
+\tilde \gamma_{ik} \beta^k_{~,j}+\tilde \gamma_{jk} \beta^k_{~,i}
-{2 \over 3}\tilde \gamma_{ij} \zero D_k \beta^k, \label{heq} \\
&&(\pa_t - \beta^k \pa_k) \tilde A_{ij} 
= {1 \over a^2 \psi^4} \biggl[ \alpha \Bigl(R_{ij} 
-{\gamma_{ij} \over 3}R \Bigr) \nonumber \\
&& \hskip 3.5cm 
-\Bigl( D_i D_j \alpha - {\gamma_{ij} \over 3}D_k D^k \alpha \Bigr)
\biggr] \nonumber \\
&& \hskip 2.5cm +\alpha (K \tilde A_{ij} - 2 \tilde A_{ik} \tilde A_j^{~k})
\nonumber \\
&& \hskip 2.5cm +\beta^k_{~,i} \tilde A_{kj}+\beta^k_{~,j} \tilde A_{ki}
-{2 \over 3} \zero D_k \beta^k \tilde A_{ij} \nonumber \\
&&\hskip 2.5cm -{8\pi\alpha \over a^2 \psi^4} 
\Bigl( S_{ij}-{\gamma_{ij} \over 3}  S_k^{~k}
\Bigr), \label{aijeq} \\
&&(\pa_t - \beta^k \pa_k) \psi + {\dot a \over 2a}\psi 
= {\psi \over 6}\Bigl\{ 
-\alpha K + \zero D_k \beta^k \Bigr\}, \label{peq} \\
&&(\pa_t - \beta^k \pa_k) K 
=\alpha(\tilde A_{ij} \tilde A^{ij}+{1 \over 3}K^2) \nonumber \\
&& \hskip 2.5cm 
-D_k D^k \alpha +4\pi \alpha (E+ S_k^{~k}), \label{keq}
\eeqn
where 
$R_{ij}$ is the Ricci tensor with respect to $\gamma_{ij}$, 
$\zero D_k$ is the covariant derivative with respect to $\eta_{ij}$, 
and 
\beq
S_{ij}=(\rho + P) u_i u_j + P \gamma_{ij}. 
\eeq

To clarify the meaning of equation (\ref{aijeq}), 
we rewrite $R_{ij}$ as 
\beq
R_{ij}=\tilde R_{ij}+R^{\psi}_{ij},
\eeq
where $\tilde R_{ij}$ is the Ricci tensor with respect to 
$\tilde \gamma_{ij}$ and 
\beqn
R^{\psi}_{ij}=&&-{2 \over \psi} \tilde D_i \tilde D_j \psi - 
{2 \over \psi} \tilde \gamma_{ij} \tilde \Delta \psi \nonumber \\
&& + {6 \over \psi^2} \tilde D_i \psi \tilde D_j \psi 
-{2 \over \psi^2} \tilde \gamma_{ij} \tilde D_k \psi 
\tilde D^k \psi. 
\eeqn
$\tilde R_{ij}$ is written as
\beqn
\tilde R_{ij}&=&{1 \over 2}\biggl[
-\zero \Delta \tilde \gamma_{ij}+\zero D_j \zero D^k \tilde 
\gamma_{ki}
+\zero D_i \zero D^k \tilde \gamma_{kj} \nonumber \\
& &~~~+2\zero D_k( f^{kl} C_{l,ij} )-2 C^l_{kj}C^k_{il}\biggr],
\label{eqij}
\eeqn
where $\zero \Delta$ is the 
Laplacian with respect to $\eta_{ij}$, 
$f^{kl}=\tilde \gamma^{kl}-\eta^{kl}$, and 
\beq
C^k_{ij}=
{\tilde \gamma^{kl} \over 2}\Bigl(\zero D_i \tilde \gamma_{jl}
+\zero D_j \tilde \gamma_{il}-\zero D_l \tilde \gamma_{ij} \Bigr).
\eeq
Note that we use a relation 
$\tilde \gamma^{ij}\zero D_k \tilde \gamma_{ij}
=\eta^{ij} \zero D_k \eta_{ij}=0$ to derive Eq. (\ref{eqij}). 
{}From Eq. (\ref{eqij}), it is found that 
under an appropriate gauge condition such as 
a transverse-traceless (TT) gauge, 
$\zero D^k \tilde \gamma_{kj}=0$, Eqs. (\ref{heq}) and 
(\ref{aijeq}) are found to constitute a wave equation for 
tensor $\tilde \gamma_{ij}$.

Hamiltonian and momentum constraint equations are 
\beqn
&& R_k^{~k}- \tilde A_{ij} \tilde A^{ij}+{2 \over 3} K^2=16\pi E,
\label{ham}\\
&& D_i \tilde A^i_{~j}-{2 \over 3}D_j K=8\pi J_j, \label{mom}
\eeqn
where 
\beqn
&& E=(\rho + P) w^2 - P,\\
&& J_i=(\rho + P) w u_i,
\eeqn
We may write the constraint equations as 
\beqn
&&\tilde \Delta \psi={\tilde R_k^{~k} \over 8}\psi - 2\pi \psi^5 a^2 E 
\nonumber \\
&& \hskip 1cm -{\psi^5 a^2\over 8}\biggl(\tilde A_{ij} \tilde A^{ij} -
{2 \over 3} K^2 \biggr),\label{ham2} \\
&& \tilde D^j (\psi^6 \tilde A_{ij}) -{2 \over 3} \psi^6 
\tilde D_i K = 8\pi J_i \psi^6,
\eeqn
where $\tilde \Delta$ is the Laplacian with respect to 
$\tilde \gamma_{ij}$. 

In this paper, we choose a constant mean curvature slice as 
we chose in a previous paper \cite{SNNM}
\beq
K=K(t)=-{3 \dot a \over a}.
\eeq
This choice can most effectively factor out overall factors 
of the expansion of the background universe from the dynamical variables. 
In this case, we obtain the equation for $\alpha$ as 
\beqn
\Delta \alpha =&& \alpha [4\pi \{ 2(\rho + P)(w^2-1)
+ \rho-\rho_0 + 3(P-P_0) \} \nonumber \\
&&~~~
+ \tilde A_{ij} \tilde A^{ij}] + 12 \pi (\rho_0 + P_0) (\alpha-1),
\eeqn
where $\Delta $ is the Laplacian with respect to $\gamma_{ij}$. 
We also note that Eqs. (\ref{peq}) and (\ref{ham2}) are, 
respectively, rewritten as 
\beqn
&&(\pa_t - \beta^k \pa_k) \psi ={\dot a \over 2a}\psi (\alpha-1)
+{\psi \over 6\eta^{1/2}} (\eta^{1/2}\beta^k)_{,k}, \\
&& \tilde \Delta \psi = {\tilde R_k^{~k} \over 8}\psi -2\pi \psi^5 a^2
[(\rho + P)(w^2 -1)+\rho-\rho_0] \nonumber \\
&&\hskip 3cm -{\psi^5 a^2 \over 8} \tilde A_{ij} \tilde A^{ij}.
\eeqn
Thus, in the constant mean curvature slice condition, 
the equations for $\alpha$ and $\psi$ are similar to those 
for the maximal slice condition, $K=0$, in the 
asymptotically flat spacetime. Hence, we expect that it 
has a singularity avoidance property for the case of 
black hole formation. 

\section{Long wavelength limit}

In this section, we consider the so-called long wavelength 
approximation assuming that the characteristic length scale $L$ of 
a density fluctuation is always much larger than 
the Hubble horizon scale, $L\gg t \sim a/\dot a$. 
First, we introduce a small parameter $\epsilon$, and 
assume that $\delta \equiv \rho/\rho_0-1$ is of $O(\epsilon^2)$ and 
its characteristic length scale is of $O(1/\ep)$. 
The latter assumption is equivalent to assuming that the magnitude of 
spatial gradients of the quantities is given by 
$\pa_i \psi = \psi \times O(\epsilon)$, 
$\pa_i \alpha = \alpha \times O(\epsilon)$, $\pa_i \delta 
=\delta \times O(\ep)$ and so on. 
Then, it is found from the equations presented in Sec. II 
that the following relation should hold:
\beqn
&& \psi-1 = O(\epsilon^0),\nonumber \\
&& u^i= O(\ep^1), \nonumber \\
&& \tilde A_{ij},~h_{ij}(\equiv \tilde \gamma_{ij}-\eta_{ij}),~
\chi(\equiv \alpha-1),~\delta = O(\epsilon^2),\nonumber \\
&& u_i,~v^i+\beta^i = O(\epsilon^3). \label{eps}
\eeqn
Here we have assumed for simplicity 
that the amplitude of primordial gravitational wave 
perturbations is negligible. We note that because the gravitational wave 
perturbations do not decay with time evolution, 
the following solution can change considerably 
if their amplitude is not small initially. 
We have also assumed that the vector part
is absent for any quantity; in other words, we do not consider
vorticity. 

It is worth mentioning that these assumptions are naturally
realized in most of successful inflation models. In the inflationary
universe scenario, only the so-called growing mode perturbations
of scalar and tensor types survive and amplitude of the tensor
perturbation is generally very small.
We note that $\psi-1$ may be larger than unity, i.e.,
we have not used any approximation for treating $\psi$. On the 
other hand, other quantities should be small enough and are regarded as 
small perturbations. If $\psi-1\ll1$ (i.e., if the linear theory
applies), the above corresponds to assuming the existence of only the
adiabatic growing mode.

Because we have not yet imposed any condition 
on $\beta^k$, we cannot specify the 
order of magnitude of $\beta^k$ and $v^k$. 
For example, in the case of 
the minimum distortion gauge\cite{smarr}, 
they are of $O(\ep)$. On the other hand, 
in the $\beta^k=0$ gauge,  $v^k=O(\ep^3)$. 
In the following discussion, we do not have to 
specify the gauge condition for $\beta^k$ in order to  
obtain solutions except for $\beta^k$ and $v^k$.

Substituting the lowest order terms in $O(\ep)$ of each variables 
shown in Eqs. (\ref{eps}), we have the following equations:
\beqn
&& {1 \over \Gamma} \dot \delta + {6\dot \psi \over \psi}
+ \nabla_k v^k=O(\ep^4),\label{per1} \\
&&  \pa_t \{ \rho_0 a^5  \nabla_k(v^k+\beta^k)\} 
+{6\dot \psi \over \psi} \rho_0 a^5 \nabla_k(v^k+\beta^k) 
\nonumber \\
&& \hskip 1.5cm 
+\rho_0a^5\nabla_k \Bigl\{{4\dot \psi \over \psi}(v^k+\beta^k)
\Bigr\} \nonumber \\
&&\hskip 1.5cm =-\rho_0 a^3 
\Bigl( \nabla^2 \chi + {\Gamma-1 \over \Gamma}
  \nabla^2 \delta \Bigr)+O(\ep^4),\label{per2} \\
&& {6\dot \psi \over \psi}-3 {\dot a \over a} \chi= 
\nabla_k \beta^k+O(\ep^4), \label{per3} \\
&& \zero \Delta \psi = -2\pi a^2 \psi^5 \rho_0 \delta+O(\ep^4),
\label{per4} 
\\
&& \nabla^2 \chi=4\pi\rho_0 a^2 \{ (3\Gamma-2) \delta 
+ 3\Gamma \chi \}+O(\ep^6),\label{per5} \\
&& \pa_t h_{ij}=-2\tilde A_{ij}
+\delta_{ik} \beta^k_{~,j}+\delta_{jk} \beta^k_{~,i}
-{2 \over 3}\delta_{ij} \beta^k_{~,k} \nonumber \\
&& \hskip 2cm +O(\ep^4),\label{per6} \\
&& \pa_t \tilde A_{ij}  + 3 {\dot a \over a} \tilde A_{ij} 
\nonumber \\
&&~~={1 \over \psi^4 a^2}\biggl[-{2 \over \psi}\Bigl(
\zero D_i \zero D_j \psi 
-{\eta_{ij} \over 3} \zero \Delta \psi \Bigr) \nonumber \\
&&~~~~ \hskip 1cm +{6 \over \psi^2}\Bigl( \zero D_i \psi \zero D_j \psi
-{\eta_{ij} \over 3} \zero D_k \psi \zero D_k \psi \Bigr) \biggr]
\nonumber \\
&&~~~~+O(\ep^4),\label{per7}
\eeqn
where 
\beqn
&& \nabla_k = {1 \over \psi^6\eta^{1/2}} \pa_k \psi^6 \eta^{1/2},
\nonumber \\
&& \nabla^2 = 
{1 \over \psi^6 \eta^{1/2}} \pa_k (\psi^2 \eta^{1/2} \eta^{kl}
\pa_l). 
\eeqn
The first two equations, (\ref{per1}) and (\ref{per2}), 
are derived from hydrodynamic 
equations and other five, (\ref{per3})$-$(\ref{per7}), 
are derived from equations for geometric variables. 

{}From Eqs. (\ref{per1}) and (\ref{per3}), we find that 
the following relations have to hold; 
\beqn
&& \dot \psi=O(\ep^2),\\
&& {1 \over \Gamma} \dot \delta+3{\dot a \over a}\chi=
-\nabla_k (v^k+\beta^k)=O(\ep^4). 
\eeqn
Also, we find that the right-hand side of Eq. (\ref{per5}) has to be 
of $O(\ep^4)$, i.e., 
\beq
(3\Gamma-2)\delta+3\Gamma \chi=O(\ep^4).
\eeq
{}From these relations and a reasonable assumption 
that $\delta \rightarrow 0$ for $t\rightarrow 0$, 
we finally find the 
time dependence for each variable at the leading order 
in $O(\ep)$ as 
\beqn
&& \delta=-{3\Gamma \over 3\Gamma-2}\chi \propto t^{2-4/3\Gamma},
\label{kai1} \\
&& \nabla_k (v^k + \beta^k) \propto t^{3-8/3\Gamma}, \\
&& u_k \propto t^{3-4/3\Gamma},\\
&& \psi \propto t^0 ,\\
&& \tilde A_{ij} \propto t^{1-4/3\Gamma}. \label{kai5}
\eeqn
Time dependence of $v^i$, $\beta^i$, $O(\ep^2)$ part of $\psi$, 
and $h_{ij}$ is found when we give spatial gauge condition for 
$\beta^i$. For example, in the TT gauge and/or 
minimum distortion gauge \cite{smarr}, 
it is found that 
\beqn
&& v^k,~\beta^k \propto t^{1-4/3\Gamma},\\
&& h_{ij} \propto t^{2-4/3\Gamma},\\
&& O(\ep^2)~ {\rm part~of~} \psi \propto t^{2-4/3\Gamma}. 
\eeqn

We emphasize again that 
we do not restrict our attention to the case 
where $\psi-1 \ll 1$. Namely, even when the scalar 
part of the metric is non-linear, 
we can still find the analytic solution as long as 
the long wavelength approximation holds. 

The purpose of this paper is to give a framework to investigate
the primordial black hole formation process, and 
its standard formation scenario is as follows: In a very 
early phase of the universe, just after inflation, the scalar-type
perturbations generated from the quantum fluctuations of an inflaton
field have the length scale much larger than the 
Hubble horizon scale at that
time. Some of these perturbations may have a large metric perturbation
amplitude\cite{BeLiWa,yokoyama,carr,ivanov}. 
As long as its scale is larger than the Hubble 
horizon scale, it never collapses, but once the scale 
becomes smaller than the Hubble 
horizon scale, it collapses to form a black hole.

The advantage of our present formalism is as follows: 
Once we give a realistic initial condition at the very 
early epoch just after inflation, evolutions of the metric and density
fluctuations can be analytically calculated as long as the length scale
of the fluctuation is much larger than the Hubble horizon scale. 
Then we may start a numerical simulation sufficiently before the scale
enters the Hubble horizon and follow the black hole formation process
without changing the gauge condition or numerical matching of initial data.
Hence, we can consistently investigate the evolution of the metric and
density fluctuations throughout the whole dynamical range starting from
a very early epoch of the universe at which analysis can be performed in
an analytical manner up to the formation of a black hole when analysis 
should be done in numerical relativity. 

\section{Numerical study in the spherical symmetric case}

To demonstrate the usefulness and robustness of our formalism,
in this section, we perform numerical simulations of primordial black
hole formation assuming spherical symmetry.

\subsection{Basic equations}

For the spherical case, 
the line element can be written as
\beqn
ds^2&&=-(\alpha^2-\psi^4\beta^2r^2 )dt^2+2\psi^4 a^2\beta rdr dt 
\nonumber \\
&& + \psi^4a^2(dr^2 + r^2 d\Omega),
\eeqn
where $\beta$ denotes $\beta^r/r$. 
We may say that we choose the minimum distortion gauge condition 
in this line element because $\pa_t \tilde \gamma_{ij}=0$ 
\cite{smarr}. As we mentioned in Sec. II, it is not always 
necessary to take 
the spatial gauge condition used here, and we may use other gauge 
conditions such as $\beta^r=0$. The spatial gauge condition 
we choose here is only one example \cite{gauge}. 

Since gravitational waves are not generated in the spherically
symmetric spacetime, we need not solve the evolution equations 
for the geometrical variables if we solve the constraint equations and
the equations of the gauge condition.
On the other hand, if we solve the evolution equations, 
we can use the constraint equations in order to check the 
accuracy of numerical solutions in each time step. 
Thus, we solve the evolution equation for $\psi$ instead 
of the Hamiltonian constraint equation, but use the latter to 
check the numerical accuracy. Then, the equations for the geometric
variables solved in numerical computation are as follows:
\beqn
&& (\pa_t - 2 \beta y \pa_y )\psi={\dot a \over 2a}\psi (\alpha-1)
+{\psi \over 6}(3\beta + 2y\pa_y \beta ),\label{spsieq}\\
&& \zero \Delta \zeta =\zeta \psi^4 a^2\Bigl[ 6\pi 
(\rho+P)(w^2 -1) \nonumber \\ 
&& \hskip 2.4cm +2\pi \{\rho-\rho_0 + 6(P-P_0)\}
\nonumber \\
&& \hskip 2.4cm +12\pi(\rho_0+P_0) + {21 \over 16}A^2 \Bigr]
\nonumber \\
&& \hskip 1.2cm +\psi^5 a^2\Bigl[ 4\pi\{2 (\rho+P)(w^2-1)
\nonumber \\
&&\hskip 2.4cm +\rho-\rho_0+3(P-P_0)\}
+{3 \over 2}A^2 \Bigr],\label{salpeq} 
\\
&& y \pa_y \beta ={3 \over 4}\alpha A,\label{sbetaeq} \\
&& \pa_r (r^3 \psi^6 A)=8\pi  r^4 \psi^6 (\rho+P)w u,\label{smoneq}
\eeqn
where $y=r^2$, $u=u_r/r$ and $\zeta=\psi(\alpha-1)$. 
We use the relation $-\tilde A_{\theta}^{~\theta}/2=
-\tilde A_{\vp}^{~\vp}/2=\tilde A_{r}^{~r} \equiv A$. 
Eq. (\ref{salpeq}) is solved under 
the boundary condition at $r \rightarrow \infty$ as 
\beq
\zeta={C \over r} e^{-r/r_0} + O(r^{-2}),
\eeq
where $C$ is a constant, and 
$r_0=1/\sqrt{12\pi\Gamma \rho_0 a^2}$. 

The Hamiltonian constraint equation 
\beqn
&& \zero \Delta \psi =  -2\pi \psi^5 a^2
[(\rho + P)(w^2 -1)+\rho-\rho_0] \nonumber \\
&&\hskip 3cm -{3\psi^5 a^2 \over 16} A^2,\label{shameq}
\eeqn
is solved only on the initial time slice with the outer boundary 
condition $\psi\rightarrow 1+ C_{\psi}/2r + O(r^{-3})$ where 
$C_{\psi}$ is a constant. 

The hydrodynamic equations are written in the form 
\beqn
&& \pa_t (w\psi^6 a^3 \rho^{1/\Gamma})
+r^{-2}\pa_{r} ( r^3 w \psi^6 a^3 \rho^{1/\Gamma} v)=0,
\label{sconeq} \\
&& \pa_{t} (w  \psi^6 a^3 (\rho +P) u)
+r^{-3}\pa_{r} ( r^4 w \psi^6 a^3 (\rho +P) v u)
\nonumber \\
&&~~~= - 2\alpha \psi^6 a^3 \pa_y P 
+w \psi^6 a^3(\rho+P)\Bigl\{
-2 w \pa_y \alpha  \nonumber \\ 
&& \hskip 2cm + u (\beta + 2y\pa_y \beta)+
{4\alpha u^2 y \over \psi^5 a^2 w} \pa_y \psi \Bigr\},
\label{hydro2}
\eeqn
where $v=v^r/r$. Using the relation, 
$\rho_0^{1/\Gamma} a^3=$constant, Eq. (\ref{sconeq}) may be written 
as
\beq
\pa_t (w\psi^6 e)
+r^{-2}\pa_{r} ( r^3 w \psi^6 e v)=0,
\eeq
where $e=(\rho/\rho_0)^{1/\Gamma}$. 
In numerical simulation ,we choose $D\equiv w\psi^6 e$ and 
$S \equiv w\psi^6 (\rho+P) u$ as variables to be solved. 
Once $D$ and $S$ are given, 
$w$ is obtained by solving the algebraic equation 
\beq
\rho_0^2\psi^{12} \Gamma^2 \biggl({D \over \psi^6} 
\biggr)^{2\Gamma}(w^2-1)={S^2y \over \psi^4 a^2}w^{2\Gamma-2}.
\eeq
and $v$ is then given from
\beqn
v=&&-\beta+{\alpha u \over w \psi^4 a^2} \nonumber \\
=&&-\beta+{\alpha S \over \Gamma w^2 \psi^{10} \rho_0 a^2 
(D/w\psi^6)^{\Gamma} }. 
\eeqn

To examine whether a black hole is formed or not, 
in each time step we search for apparent horizon 
which is defined as the outermost trapped surface \cite{hawking}.  
In the spherically symmetric case, 
the outermost zero point, $r=r_{\rm AH}$, of the function 
\beq
\Theta(r)\equiv 2 {\dot a \over a}+A + {1 \over \psi^2 a} 
\biggl( {2 \over r}+{4 \pa_r \psi \over \psi}\biggr),
\eeq
corresponds to the outermost trapped surface \cite{SNNM}. 
Hence, we only need to calculate $\Theta$ in each time step 
and look for the zero point. 

Once the apparent horizon is determined, we also calculate the 
mass of the apparent horizon which we define as 
\beq
M_{\rm AH}={a\psi^2 r \over 2} ~{\rm at}~r=r_{\rm AH}.
\eeq
$M_{\rm AH}$ is not identical with the gravitational mass of a 
black hole in general. However, if it settles down to a constant 
in the late epoch after formation of the black hole, it 
can be regarded as the gravitational mass because 
in the spherical and static space $M_{\rm AH}$ agrees with the 
gravitational mass. 

In order to check the numerical accuracy of $M_{\rm AH}$, 
we also calculate the conserved Kodama mass in the spherical 
spacetime \cite{Kodama}. The Kodama mass within 
a radius $r$ is defined as 
\beq
M_{\rm K}(r)
=4\pi\int_0^r r'^2 dr' a^3 \alpha \psi^6 T^t_{~\mu} K^{\mu},
\eeq
where the components of $K^{\mu}$ are 
\beqn
&&K^t=-{1 \over \alpha \psi^2} {\pa \over \pa r}\psi^2 r ,\\
&&K^r={1 \over \alpha \psi^2} {\pa  \over \pa t}\psi^2 r,
\eeqn
and $K^{\theta}=K^{\varphi}=0$. Since 
$M_{\rm K}$ at $r=r_{\rm AH}$ is proved to be equal to 
$M_{\rm AH}$ \cite{Kodama}, it can be used to check the accuracy 
of our estimation of $M_{\rm AH}$. In our simulations, 
we found that both agree well except for the very late 
epoch after formation of a black hole at which 
the gradient of $\alpha$ near the apparent horizon is very 
steep and it is difficult to keep numerical accuracy well. 

Numerical simulations are performed taking $3000$ inhomogeneous 
grid points for the $r$-axis. The circumferential radius of 
the outer boundary is always kept to be much larger than the 
Hubble horizon
scale. We also take a sufficient number of grids inside a black hole 
forming region. More concretely, we take grids 
as $r_i=\Delta r (f^i-1)/(f-1)$, where $i=0,1,2,\cdots 3000$, 
$f$ is a constant slightly larger than $1$, 
and $\Delta r$ is the grid spacing at origin which is much 
smaller than the circumferential length of a formed black hole. 

\subsection{Initial conditions}

We make use of the analytic solution derived in 
Sec. III to give a realistic initial condition for 
$\rho$, $u$ and so on. 
Thus, the initial condition is given 
at an epoch when the length scale of a density fluctuation is 
much larger than the Hubble horizon scale. 

First, we assume that 
$\delta=\rho/\rho_0-1$ is much smaller than unity and write it as 
\beq
\delta = f(r) t^{2-4/3\Gamma}.
\eeq
{}From Eq. (\ref{kai1}), we soon get $\chi=\alpha-1$ as 
\beq
\chi=-{3\Gamma-2 \over 3\Gamma} f(r) t^{2-4/3\Gamma}.
\eeq
$u$ is derived from Eq.~(\ref{hydro2}) in the long 
wavelength limit as 
\beq
u={2 \over 3\Gamma+2} \pa_y f(r) t^{3-4/3\Gamma}.\label{uinit}
\eeq
Thus, if we specify the function $f(r)$ on the initial time slice 
$t=t_0$, and subsequently solve Eqs. 
(\ref{salpeq})$-$(\ref{smoneq}) and (\ref{shameq}), 
we obtain the initial data for $A$, $\beta$, $v$, $\psi$ and 
$\alpha$. 

In this paper, we simply give $\delta$ as 
\beqn
\delta \cdot \psi^6 =&& C_{\delta} \biggl[ 
\exp\Bigl(-{r^2 \over r_0^2 }\Bigr) -\sigma^{-3} 
\exp\Bigl(-{r^2 \over \sigma^2r_0^2 }\Bigr)\biggr]\nonumber \\
&&\hskip 1cm \times \biggl({t \over r_0}\biggr)^{2-4/3\Gamma}
,\label{delpsi}
\eeqn
where $C_{\delta}$, $r_0$ and $\sigma$ 
are constants, and we set $a_f=r_0^{-2/3\Gamma}$. 
Roughly speaking, 
$C_{\delta}$ and $\sigma$ specify the amplitude and 
shape of the density fluctuation, respectively. 
$r_0$ determines the length scale of the density fluctuation, 
and we fix it to be unity in the following. Hence, hereafter, 
the mass and length are shown in the units $r_0=1$. 

If we define the spectrum of density fluctuations as 
\beq
\delta(k)\equiv \int_0^{\infty} j_0(kr) \delta \cdot \psi^6r^2dr,
\eeq
where $j_0(kr)\equiv \sin(kr)/kr$ is the spherical Bessel function of
the 0-th order, we get
\beqn
\delta(k)&&={\sqrt{\pi}r_0^3 \over 4}
\biggl({t \over r_0}\biggr)^{2-4/3\Gamma} C_{\delta} \nonumber \\
&&\times \biggl[
\exp\Bigl({-k^2r_0^2 \over 4} \Bigr) 
-\exp\Bigl({-k^2r_0^2\sigma^2 \over 4} \Bigr) 
\biggr].
\eeqn
Thus the wavelength of the dominant spectral components of 
the density fluctuation is larger than $\sim \pi r_0$ in the comoving
scale (or $k \alt 2 /r_0$). 
Hence, we start all the simulations at an initial time $t$ 
which satisfies the condition $t \ll \pi(1-2/3\Gamma) a(t)r_0$.
In the following, we always set the initial time as $10^{-4}r_0$. 

Initial conditions are numerically 
determined by performing iteration as 
follows: (a) We solve Eq.~(\ref{shameq}) for $\psi$, for the 
density profile given by Eq. (\ref{delpsi}). (b) From 
$\delta \cdot \psi^6$ as well as $\psi$ obtained in (a), 
we determine $f(r)$. 
(c) we calculate $u$, $w$, $A$, and $\beta$ by using Eqs. 
(\ref{uinit}), $w=\sqrt{1+u^2r^2\psi^{-4}}$, 
(\ref{smoneq}) and (\ref{sbetaeq}), respectively, 
and substitute the new $w$ and $A$ into Eq. (\ref{shameq}). 
We repeat this procedure until sufficient convergence 
is achieved. 

Hereafter, we pay attention only to the $\Gamma=4/3$ case, 
because it is most probable that the universe was radiation-dominated
in the early times.
In this case, for $t \rightarrow 0$, the metric behaves as 
\beqn
&&\chi  \rightarrow 0,\\
&&\beta \rightarrow {\rm const} \ll 1,\\
&&\psi-1  \rightarrow {\rm const} = O(1). 
\eeqn
Apparently, the metric is non-linear 
for large $C_{\delta}$ at $t=0$ because $\psi-1 = O(1)$. 
Note that $\delta \rightarrow 0$, for $t \rightarrow 0$. 
Hence, in the present coordinate condition, a black hole is 
formed only if the metric is non-linear even though the density 
fluctuation is very small. In other word, we may say 
that the criterion of formation of black holes depends only on 
$\psi$ at $t=0$. 

\subsection{Numerical results}

Numerical simulations were performed for various values of $C_{\delta}$
and $\sigma$. Specifically, we surveyed a two-dimensional 
parameter space with $C_{\delta}>0$ and $\sigma>1$. 
We note that in the case $\sigma=\infty$, 
the high density peak is surrounded by a flat universe, and 
in other cases, it is surrounded by a region in which the 
density is lower than that of the flat universe. 

We show the spectrum of $\delta(k)$ 
for the cases $\sigma=1.5$, 2, 3, 5, 8 and $\infty$ in Fig.~1. 
For $\sigma=\infty$, the peak is at $k=0$, while in the other 
cases, the peak wavenumber and the width around 
the peak becomes larger and wider, respectively, 
for smaller $\sigma$. 

In Fig.~2, we show $\psi_0=\psi(r=0)$ at 
$t=0$ as a function of $C_{\delta}$ for $\sigma=\infty$, 2, 3 and 5.
Here, we can easily calculate $\psi_0$ at $t=0$ by using the analytic 
solution found in Sec. III. It is found that $\psi_0$ at $t=0$ 
monotonically increases with increasing 
$C_{\delta}$, irrespective of $\sigma$. 

In Figs.~3 and 4, we show general features of 
numerical solutions taking the case $\sigma=\infty$ 
as an example. The features shown in Figs.~3 and 4 are found 
also in the cases of finite $\sigma$. 
In Fig.~3, we show 
$\delta$ and $2(1-\alpha)$ at origin as 
a function of time $t$ for black hole formation case 
($C_{\delta}=15$, dotted lines)  
and no formation case ($C_{\delta}=13$, solid lines), 
respectively. The reason why 
we plot $2(1-\alpha)$ for $\alpha$ is that it must coincide 
with $\delta$ for $t \rightarrow 0$.
Fig.~3 clearly shows that for $t \ll r_0$, 
(a) $\delta$ and $\alpha-1$ 
are proportional to $t$, and (b) $\delta$ agrees with 
$2(1-\alpha)$. Hence, the numerical solution reproduces 
the analytic solution for $t \ll r_0$ accurately. 
It is found that the difference between $\delta$ and $2(1-\alpha)$ 
gradually becomes appreciable at $t/r_0 \simeq 0.1$ and 
becomes as large as unity when $t/r_0 \sim 1$. This is 
reasonable because the density fluctuation amplitude can become
non-linear only after its length scale is smaller than the 
Hubble horizon scale.

In Fig.~ 4, we also show $\psi_0$ as a function of $t$. 
As we found in Sec. III, it does not change so much for 
$t < r_0$, but slightly decreases with time evolution. 
This is due to the effect of the $O(\ep^2)$ part in $\psi_0$. 
In some previous works, $\psi_0$ at $t \sim r_0$ is assumed 
to be equal to $\psi_0$ at $t=0$. But as shown here, 
the assumption is strictly speaking not correct. 

In Figs. 5, we show the mass of the apparent horizon ($M_{\rm AH}/r_0$) 
as a function of time ($t/r_0$) in the black hole formation cases 
for $\sigma=2$, 3, 5 and $\infty$. 
We note that in all the simulations, computation is terminated when it
is difficult to keep the numerical accuracy near the apparent horizon. 
Although the simulations had to be ended before we could draw a definite
conclusion, our results given in the figures strongly suggests that
$M_{\rm AH}$ approaches an asymptotic value without increasing forever.
This is consistent with a previous result 
\cite{bick,NieJed}. (We note that by comparing it with $M_{\rm Kodama}$,
it is found that $M_{\rm AH}$ shown here is accurate 
to within a few percent. ) 

The formation epoch of a black hole is highly 
dependent on the initial density profile. In the case $\sigma=\infty$, 
the formation epoch is $t > 100r_0$, but in the other cases, it is 
earlier for smaller $\sigma$. This is natural because for smaller 
$\sigma$, the wavenumber of the peak in the spectrum 
is larger. 
It is found that $M_{\rm AH}$ can be $> 8r_0$ 
in the case $\sigma=\infty$, 
but it is at most $\sim 1.1 r_0$ in the case $\sigma=2$. These 
facts also indicate that 
the formation epoch and the maximum mass of black holes 
are highly dependent on the initial density profile. 
We also note that even when we pay attention only to 
a particular spectrum shape, $M_{\rm AH}$ 
can vary in a large range depending on $C_{\delta}$ 
(or $\psi_0$ at $t=0$). 
Since we do not pursue the simulation 
in which a black hole in the limit $M_{\rm AH}=0$ is formed, 
we cannot draw a strong conclusion, but 
it seems possible to have a black hole whose mass is much smaller than 
$r_0$ near the threshold of black hole formation. 

In Figs.~6, we plot $M_{\rm AH}$ as a function of $\psi_0$ at $t=0$ 
for the cases $\sigma=2$, 5 and $\infty$ as examples. 
We find that 
the threshold of $\psi_0$ at $t=0$ for formation of black hole is 
quite different among all the models. 
In the case $\sigma=\infty$, 
the threshold value is high ($\psi_0(t=0) \sim 1.79$). 
On the other hand, it is not so large 
in other cases, and smaller for smaller $\sigma$.
In Fig.~7, we plot the threshold line on the
($\sigma,\psi_0(t=0)$)-plane. 
The reason of this property is simple: In the case in which 
the density peak is surrounded by a high density region, 
it is forced to be expanded by the 
surrounding region more strongly and is prevented from 
collapsing. As a consequence, the density peak 
surrounded by a higher density region is 
more difficult to form a black hole, while 
the density peak surrounded by a lower density region can 
more easily form a black hole. Probably, 
the density peak surrounded by a zero density region can 
most easily form a black hole. 
Motivated by this observation, we also perform simulations 
in the flat spacetime from the time symmetric 
initial condition with the initial density fluctuation profile,
\beq
\delta \cdot \psi^6=C_{\delta}\exp\Bigl(-{r^2 \over r_0^2}\Bigr). 
\eeq 
In this case, the threshold value for $\psi_0$ at $t=0$ 
is about $1.428$. 
This value can be approximately regarded as the smallest value of 
$\psi_0$ at $t=0$ for the formation of a black hole. 

As another useful criterion, we point out an approximate 
measure for determining the formation of black holes in the special case 
of spherical symmetry. 
First, by using the Kodama mass, we define an excess 
mass at each radius at $t=0$ as 
\beqn
\delta M_{\rm K}(r) && \equiv M_{\rm K}-M_{\rm F} \nonumber \\
&&= 4\pi \rho_0 a^3 \int_0^r r'^2 dr' \delta \cdot \psi^6 
\biggl(1+{2r' \over \psi}{d\psi \over dr}\biggr),
\eeqn
where $M_{\rm F}$ denotes the mass in the case of 
a non-perturbed universe and we have used the relations which hold at
$t=0$ such as $u=O(t^2)$, $\alpha=1+O(t)$, and $\pa_t \psi=O(t)$ to
neglect the terms higher order in $t$. 
Then, we define a compaction function at each radius as 
\beq
C(r)={\delta M_{\rm K} \over r \psi(r)^2 a}. \label{comp}
\eeq
Here, we note that this compaction can be defined at $t=0$, because 
it approaches a time-independent function in the limit 
$t \rightarrow0$.
In Fig.~8, we show (\ref{comp}) as a function of $r$ for 
some of filled circles on the thick dotted line in Fig.~7.
It is found that the maximum value $C(r)_{\rm max}$ 
is about $0.4$ irrespective 
of $\sigma$. We note that for each $\sigma$, 
if the maximum value is large than the value shown in Fig.~8, 
a black hole is always formed, while if not, it is not. 
Thus, the measure presented here will be helpful as an 
approximate criterion to know whether a spherical black hole is 
formed only from the initial data at $t=0$, and 
if we use $C(r)_{\rm max}$ as a parameter instead of $\psi_0(t=0)$, 
we can approximately neglect the dependence on $\sigma$. 

\section{Summary}

We have presented a formulation for numerical relativity in the
cosmological background by which we can perform a numerical simulation
of primordial black hole formation with the initial data that can be
analytically given. Namely, we have formulated the Einstein and
hydrodynamic equations in the constant mean curvature time-slicing in a
way suitable both for obtaining the analytic solution of a perturbation
while its length scale is well over the Hubble horizon scale and for
performing a numerical simulation until the formation of a black hole.
As a result, it becomes possible to investigate the 
primordial black hole formation from a very early phase of the universe
just after inflation up to the formation of black holes in a continuous
manner, without changing the gauge condition or numerical matching of
the initial data.

By using our formalism, we have carried out a numerical simulation of
the black hole formation in a spherically symmetric, 
radiation-dominated universe, starting from a realistic initial 
data which are given by the analytical solution of a 
superhorizon scale perturbation. 
In this paper, we have considered the initial conditions which are 
specified by two parameters; one characterizes the amplitude of 
the density (or metric) fluctuation and the other 
the shape of the density profile. 
It is found that the formation criterion is moderately dependent on 
the initial profile of the density fluctuation. In the case when 
the density peak is surrounded by a flat Friedmann universe, 
the threshold value of $\psi_0$ at $t=0$ for forming a black hole 
is very large $\sim 1.8$, while when surrounded by a 
low density region, it may be as small as $\sim 1.4$. 
This property suggests that the formation of primordial 
black holes may not be determined by a local criterion: 
Even when there is a density fluctuation of a high density contrast,
it may not efficiently collapse into a black hole if it is 
in a high density region, but it efficiently collapses if it is 
in a low density region. 
As we have noted, this moderate variation of the formation criterion 
is translated to a very large variation in the actual number of
primordial black holes.
Thus, we conclude that the spatial correlation of primordial density
fluctuations is crucially important in studying 
the formation of primordial black holes.

In this paper, we have assumed the spherical symmetry and restricted 
our attention to a model which contains only two parameters. Even in
this case, the formation criterion was not so simple. 
In reality, a primordial black hole is formed in a spacetime 
without any spatial symmetries. In such a case, the 
anisotropic effect will be important in addition to the inhomogeneous
effect shown in this paper, and it is expected that 
the formation criterion will be much more complicated than the present
case. Apparently, the next step is to carry 
out non-spherical simulations, and even for such a simulation, the 
formalism presented here is perfectly applicable. 

\acknowledgments

We thank K. Nakao for useful discussion 
and for giving us a note in which
Kodama's mass is concisely described. 
M. Shibata thanks Stu Shapiro for helpful comments. 
This work was supported in 
part by the Monbusho Grant-in-Aid for scientific research 
Nos.~08NP0801, 09640355 and 09740336. M. Shibata's work was also 
supported in part by the JSPS Fellowships for Research Abroad.

\clearpage

\begin{figure}[t]
\epsfxsize=3in
\leavevmode
\epsffile{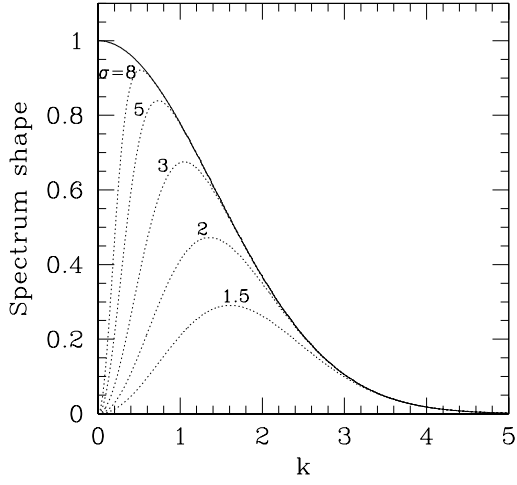}
\caption{The spectrum shapes of the density fluctuation 
($\exp(-k^2/4)-\exp(-k^2\sigma^2/4)$) are shown 
for several $\sigma$. 
The solid line denotes the case for $\sigma=\infty$ 
in which the density peak is surrounded by a 
flat universe, and 
the dotted lines denote the cases for $1.5 \leq \sigma \leq 8$. }
\end{figure}

\begin{figure}[t]
\epsfxsize=3in
\leavevmode
\epsffile{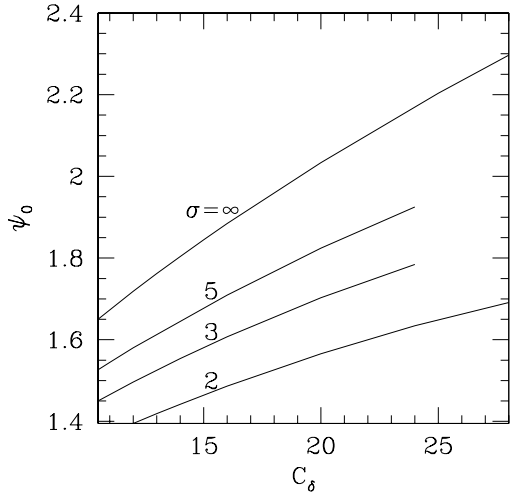}
\caption{$\psi_0$ at $t=0$ as a function of $C_{\delta}$ 
for $\sigma=\infty$, 2, 3 and 5.  
}
\end{figure}

\begin{figure}[h]
\epsfxsize=3in
\leavevmode
\epsffile{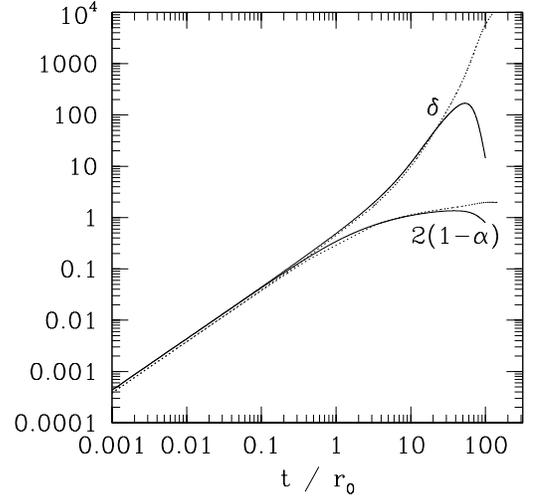}
\caption{$\delta$ and $2(1-\alpha)$ at origin as 
a function of time $t$ for black hole formation case 
($C_{\delta}=15$, dotted lines)  
and no formation case ($C_{\delta}=13$, solid lines), 
respectively. 
}
\end{figure}

\begin{figure}[h]
\epsfxsize=3in
\leavevmode
\epsffile{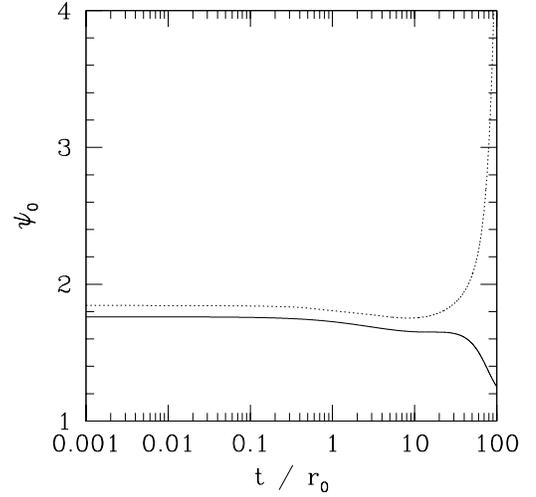}
\caption{The same as Fig. 3, but for $\psi_0$ as 
a function of time $t$.   
}
\end{figure}

\clearpage

\begin{figure}[h]
\epsfxsize=3in
\leavevmode
(a)\epsffile{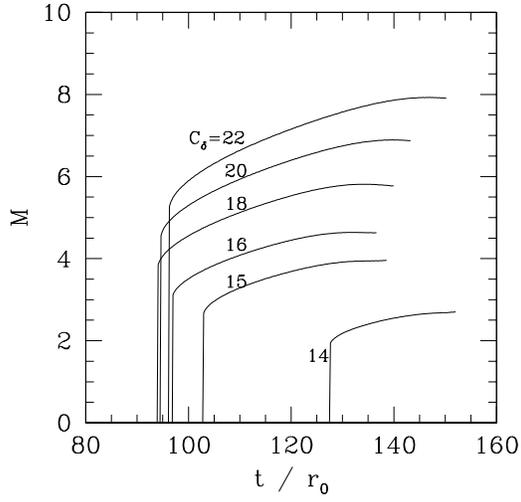}
\epsfxsize=3in
\leavevmode
(b)\epsffile{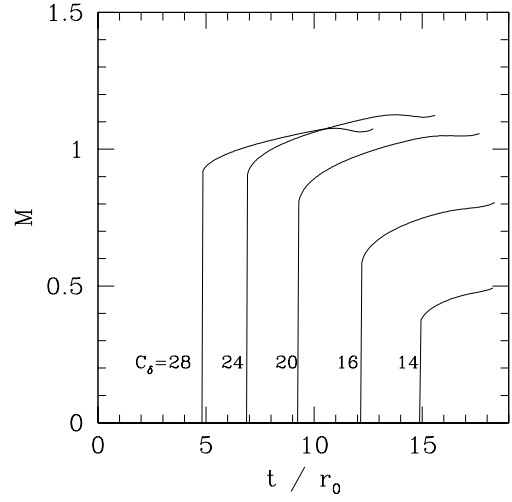}\\
\epsfxsize=3in
\leavevmode
(c)\epsffile{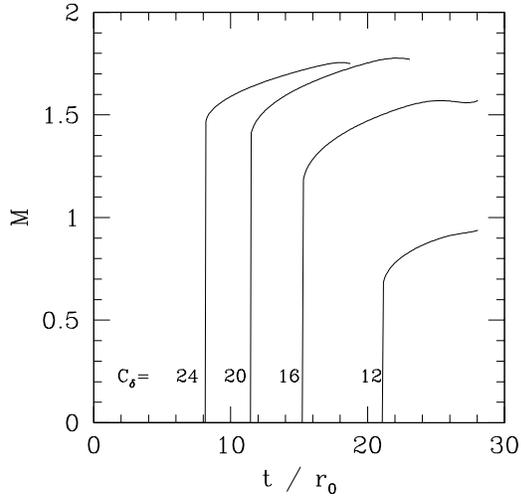}
\epsfxsize=3in
\leavevmode
(d)\epsffile{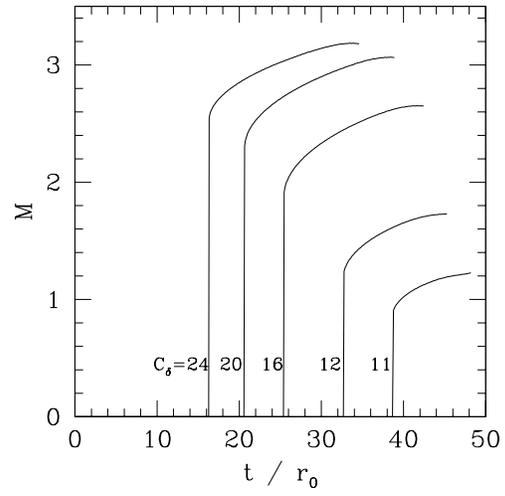}
\caption{$M_{\rm AH}/r_0$ as 
a function of time $t/r_0$ in black hole formation cases 
for $\sigma=\infty$(a), 2(b), 3(c), and 5(d), respectively. 
}
\end{figure}

\clearpage

\begin{figure}[h]
\epsfxsize=3in
\leavevmode
\epsffile{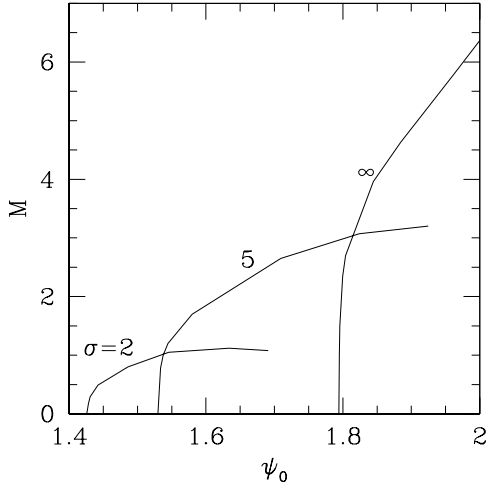}
\caption{$M_{\rm AH}/r_0$ as a function of $\psi_0$ at $t=0$ for 
$\sigma=\infty$, 2 and 5. 
}
\end{figure}

\begin{figure}[h]
\epsfxsize=3in
\leavevmode
\epsffile{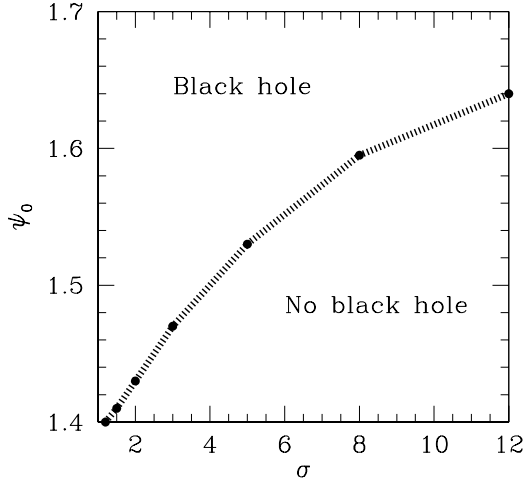}
\caption{Summary of numerical results on formation 
of black holes in $\sigma-\psi_0(t=0)$ plane. For 
initial conditions located in the region 
above the thick dotted line, we find that a black hole is formed. 
Note that the solid line reaches $\psi_0(t=0) \simeq 1.79$ in the limit 
$\sigma \rightarrow \infty$. 
}
\end{figure}

\begin{figure}[h]
\epsfxsize=3in
\leavevmode
\epsffile{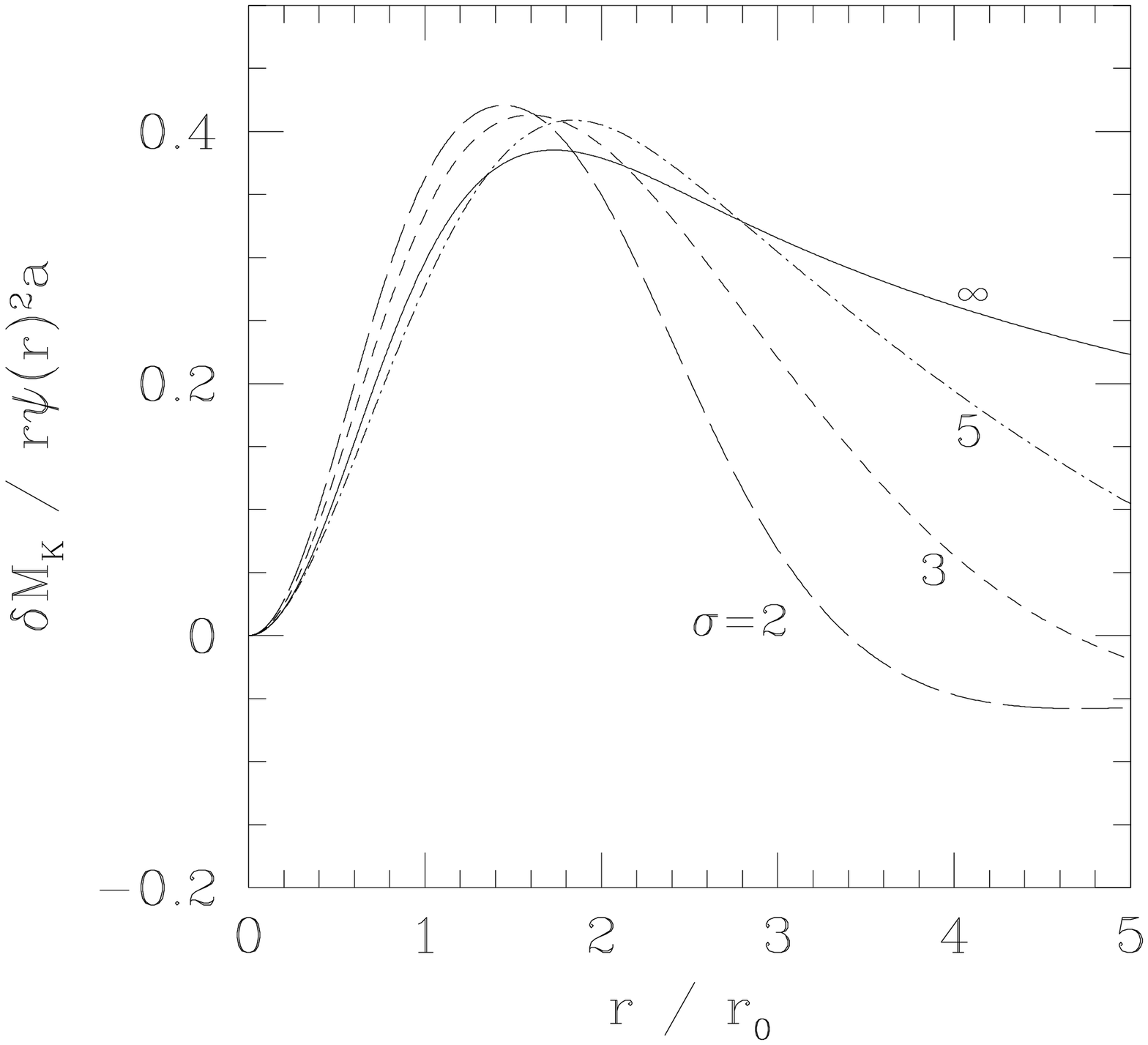}
\caption{$\delta M_{\rm K}/r\psi(r)^2a$ as a function of 
$r$ at $t=0$ for the critical cases of $\sigma=2$, 3, 5 and 
$\infty$ (i.e., for the initial condition 
denoted by filled circles in Fig. 7). 
The maximum value is $\sim 0.4$ irrespective of $\sigma$. 
If the maximum value $C(r)_{\rm max}$ 
is larger than $\sim 0.4$, a black hole is formed, but if not, 
it is not for all $\sigma$.
}
\end{figure}


\begin{thebibliography}{99}
\bibitem{gauge} In particular, for 3D simulation in which 
the grid number we can take is restricted, 
the minimum distortion gauge 
may not be appropriate for a black hole formation problem, 
because the coordinate spreads outward (i.e., $\beta^i > 0$) 
when a matter is collapsing (i.e., $v^i < 0$) 
and hence it is soon difficult to keep 
the numerical accuracy around the collapsing matter. 
In the case of 3D simulation, it will be necessary to choose 
the spatial gauge condition more carefully.
\end{thebibliography}
\end{document}